\documentclass[a4paper,12pt]{article}

\usepackage{latexsym}
\usepackage{amsmath}
\usepackage{amssymb}
\usepackage{enumitem}

\begin{document}

\title{On the ``Legendre scalarization'' of nonlinear gravity theories}

\author{Guido MAGNANO \\
\small Dipartimento di Matematica ``G.~Peano'',
Universit\`a di Torino, \\
\small via Carlo Alberto 10, 10123 Torino, Italy}
\date{}
\maketitle

\begin{abstract}
We discuss the proposal of a new method to transform a $f(R)$ metric gravity theory into a general relativistic theory including an auxiliary scalar field, recently introduced by S.~Cotsakis et al. We argue that (i) the fact that the fourth order equations of $f(R)$ metric gravity can be recast (via a Legendre transformation) into Einstein equations \emph{without any conformal rescaling} has been thoroughly clarified in the previous literature, and (ii) the newly proposed method produces a set of equations that are not equivalent to the original theory. In the conclusion, a comment is added on another aspect of the Legendre transformation which seems to be generally overlooked.
\end{abstract}

In the article \cite{CMM} the authors present a ``Legendre scalarization'' method which is expected, starting form a metric nonlinear Lagrangian $f(R)$, to produce \emph{``a theory with second order field equations that describes general relativity with a self-interacting scalar field, without requiring the introduction of conformal frames.''}

It has been known for a long time \cite{old} that the fourth order equations produced by a Lagrangian of the form 
\begin{equation}
L=f(R)\sqrt{|g|}+L_{mat}
\end{equation}
($R=g^{\mu\nu}R_{\mu\nu}$ being the curvature scalar of the spacetime metric, $g$ being the determinant of the metric and $L_{mat}$ being the matter Lagrangian), i.e.
\begin{equation}
f'(R)R_{\mu\nu}-\frac{1}{2}f(R)g_{\mu\nu}-\nabla_\mu\nabla_\nu f'(R)+g_{\mu\nu}\Box f'(R)= T_{\mu\nu},\label{NLeq}
\end{equation}
are dynamically equivalent to Einstein's equation for a new metric defined by
\begin{equation}
\tilde{g}_{\mu\nu}=f'(R)g_{\mu\nu}\label{cr}
\end{equation}
in interaction with an additional scalar field which, on the solutions of the dynamical equations, equals $f'(R)$. Such conformal rescaling of the metric is ubiquitously referred to as ``passing to the \emph{Einstein frame}''. This originated the belief that a conformal rescaling is necessary to recast the fourth order equation (\ref{NLeq}) into a second order system including an Einstein equation. 

A known issue related with the conformal rescaling is that it produces a direct coupling of the matter Lagrangian $L_{mat}$ with the auxiliary scalar field \cite{Brans}. In cosmology, assuming that the Einstein frame metric is the spacetime metric produces different physical outcomes w.r.~to the original ``Jordan frame'' \cite{Cap}.

Thus, it is indeed reasonable to seek an alternative procedure to recast (\ref{NLeq}) into the equations for a general-relativistic self-gravitating scalar field \emph{without rescaling the metric}. 

The application of a generalised Legendre transformation to relativistic gravity theories was instead envisaged by J.~Kijowski in 1977 \cite{K1}. Roughly speaking, one identifies the \emph{configuration} of the gravitational field with the connection (which defines the geodesis worldlines), and the Ricci tensor which appears in the Lagrangian plays the role of the \emph{velocity}. The derivative of the Lagrangian w.r.~to the (symmetrized) Ricci tensor, $\pi^{\mu\nu}=\dfrac{\partial L}{\partial R_{(\mu\nu)}}$, becomes therefore the \emph{conjugate momentum} to the connection (these concepts apply also to gravity theories where the connection is non-metric and the Ricci tensor may thus be nonsymmetric). From the tensor density $\pi^{\mu\nu}$ one can easily produce the symmetric tensor $\tilde{g}_{\mu\nu}$, defined by the relation $\pi^{\mu\nu}=\tilde{g}^{\mu\nu}\sqrt{|\tilde{g}|}$. This idea was first exploited in purely affine theories, where the only dynamical variable representing gravity is a connection, and the Lagrangian is a function of its Ricci tensor (e.g.~the square root of its determinant): since there is no metric among the field variables, it is the conjugate field $\tilde{g}_{\mu\nu}$ that provides the spacetime metric.

The method has been subsequently applied also to purely metric theories with a Lagrangian $L=f(R_{\mu\nu})\sqrt{|g|}$ \cite{MFF1,JK}. In this case, the Legendre transformation produces an equivalent model in which the dynamical variables are the original metric $g_{\mu\nu}$ and the new metric $\tilde{g}_{\mu\nu}$; some independent physical assumption is needed to identify the spacetime metric with either the original or the new metric. 

In the particular case where $L=f(R)\sqrt{-g}$, one finds (in dimension four)
\begin{equation}
\tilde{g}^{\mu\nu}\sqrt{|\tilde{g}|}=\dfrac{\partial L}{\partial R_{(\mu\nu)}}=f'(R)g^{\mu\nu}\sqrt{-g}\quad\Rightarrow\quad\tilde{g}_{\mu\nu}=f'(R)g_{\mu\nu},\label{cr2}
\end{equation}
thus one may believe that in this case the Legendre transformation coincides with the conformal rescaling of the metric. However, this is not the case. 

Let us recall the exact steps of the Legendre transformation. For ease of reference, we will first review the familiar procedure of classical mechanics. The first step is to define the map linking the conjugate momenta to the dynamical state of the system, represented by the lagrangian coordinates and velocities $(q^\lambda,\dot{q}^\lambda)$:
\begin{equation}
    p_\lambda=\dfrac{\partial L}{\partial \dot{q}^\lambda}.\label{lmap}
\end{equation}
\noindent Then, provided $\det\left|\dfrac{\partial^2 L}{\partial \dot{q}^\lambda\dot{q}^\mu}\right|\ne 0$, one should invert the map above to express the velocities as functions of coordinates and momenta:
\begin{equation}
    \dot{q}^\lambda=u^\lambda(q^\mu,p_\mu) ;\label{ilmap}
\end{equation}
\noindent third, one introduces the Legendre transform of the Lagrangian,
\begin{equation}
    H(q^\mu,p_\mu)=p_\mu u^\mu - L(q^\mu, u^\mu)\label{ham}
\end{equation}
(for smooth functions, the alternative definition 
$$H(q^\mu,p_\mu)=\sup_{u^\mu\in\mathbb{R}^n}(p_\mu u^\mu - L(q^\mu, u^\mu))$$ is completely equivalent to the steps above). 
The 1-form $Ldt$ which appears in the original action integral is replaced, in the phase space, by the Poincaré-Cartan form $\theta_H=p_\mu dq^\mu - H(q^\mu,p_\mu)dt$. The pullback of $\theta_H$ on a generic curve in the phase space defines the new action integral:
\begin{equation}
    \int \big[p_\mu \dot{q}^\mu - H(q^\mu,p_\mu)\big]dt = \int \big[p_\mu (\dot{q}^\mu-u^\mu) + L(q^\mu, u^\mu)\big]dt.
\end{equation}
Therefore, at the level of action integrals the result of the Legendre transformation can be described as the replacement of the original Lagrangian $L$ with the \emph{Helmholtz Lagrangian} \cite{He}
\begin{equation}
    L_H=p_\mu (\dot{q}^\mu-u^\mu) + L(q^\mu, u^\mu),\label{Hl}
\end{equation}
whereby $u^\mu$ is the function of $(q^\mu,p_\mu)$ defined by (\ref{ilmap}). The variation of the Helmholtz Lagrangian yields a first order system of equations (the Hamilton equations); by eliminating of the momentum variables $p_\mu$ from the system -- which is always possible, since the variation of $L_H$ w.r.~to $p_\mu$ reproduces eq.~(\ref{ilmap}) -- one recovers the original (second order) Lagrange equations.

Now, let us revert to purely metric gravity theories \cite{MFF2,MFF3,M}. In the general case $L=f(R_{\mu\nu})\sqrt{|g|}$, once introduced the ``conjugate metric'' $\tilde{g}_{\mu\nu}$ one should find the inverse map, in order to express the Ricci tensor $R_{\mu\nu}$ as a function of $g_{\mu\nu}$ and $\tilde{g}_{\mu\nu}$, $R_{\mu\nu}=r_{\mu\nu}(g_{\alpha\beta},\tilde{g}_{\alpha\beta})$: this step corresponds to (\ref{ilmap}). The Legendre transform of $L$ is 
\begin{equation}
    H(q^\mu,p_\mu)=\tilde{g}^{\mu\nu}r_{\mu\nu}\sqrt{|\tilde{g}|} - f(r_{\mu\nu})\sqrt{|g|}
\end{equation}
(not to be confused with the Hamiltonian in the ADM sense), and the Helmholtz Lagrangian is
\begin{equation}
    L_H=\tilde{g}^{\mu\nu}(R_{\mu\nu}-r_{\mu\nu})\sqrt{|\tilde{g}|} + f(r_{\mu\nu})\sqrt{|g|}.\label{LH1}
\end{equation}
It turns out that the term $\tilde{g}^{\mu\nu}R_{\mu\nu}$ can be rewritten as the scalar curvature $\tilde{R}= \tilde{g}^{\mu\nu}\tilde{R}_{\mu\nu}$ plus a quadratic combination of the covariant derivatives of $g_{\mu\nu}$ (w.r.~to the Levi-Civita connection of $\tilde{g}_{\mu\nu}$), plus a full divergence that has no effect on the field equations. Thus, $L_H$ can be rewritten as the usual Einstein-Hilbert Lagrangian of GR for the new metric $\tilde{g}_{\mu\nu}$, in interaction with the other tensor field $g_{\mu\nu}$, a manifestation of the property that J.~Kijowski called \emph{universality of Einstein equations} \cite{K2}.

Although the Einstein-Hilbert term in the Lagrangian can be written for the metric $\tilde{g}_{\mu\nu}$, and not for the original metric $g_{\mu\nu}$, in the full (second order) system of equations obtained by the variation of $L_H$ w.t.~to the two \emph{independent} field variables $\tilde{g}_{\mu\nu}$ and $g_{\mu\nu}$ a few manipulations allow to obtain an Einstein equation for the original metric $g_{\mu\nu}$ as well. In other terms, once the Legendre transformation has allowed one to recast the original fourth order equations into a second order system (by doubling the independent variables), either of the two metrics can be seen as fulfilling Einstein equations, with the other metric acting as a source \cite{MS2}.

This procedure, however, is viable only if the definition of the new metric as a function of $R_{\mu\nu}$ can be inverted in order to obtain a function $r_{\mu\nu}(g_{\alpha\beta},\tilde{g}_{\alpha\beta})$. This is evidently not the case for Lagrangians of the form $f(R)$. In this case $\tilde{g}^{\mu\nu}$ and $g^{\mu\nu}$ are conformally related, as we have already seen: thus the components of $\tilde{g}^{\mu\nu}$ cannot play the role of independent ``conjugate momenta''. Therefore, it is necessary to perform a different Legendre transformation (a formal geometric description of the generalised Legendre transformation, covering all these cases, can be found in \cite{MFF3}). 

Since the Lagrangian depends on the second derivatives of the metric only through the curvature scalar $R$, the conjugate variable is now the scalar density defined by
\begin{equation}
    \pi = \dfrac{\partial L}{\partial R}=f'(R)\sqrt{|g|},\label{lmapd}
\end{equation}
to which one associates the scalar field $p$ such that $\pi=p\sqrt{|g|}$, i.e.
\begin{equation}
    p = f'(R).\label{lmap2}
\end{equation}
If $f''(R)\ne 0$, the map (\ref{lmap2}) can be inverted, i.e.~a function $r(p)$ can be defined such that $f'(r(p))\equiv p$. The Legendre transform of $L$ is 
\begin{equation}
    H=\big(p\cdot r(p) - f(r)\big)\sqrt{|g|}\label{H2}
\end{equation}
(here and in the sequel, $f(r(p))$ is abbreviated to $f(r)$ for better readability) and the Helmholtz Lagrangian is
\begin{equation}
    L_H=p(R-r(p))\sqrt{|g|} - f(r)\sqrt{|g|}=\big(p\,R + V(p)\big)\sqrt{|g|},\label{LH2}
\end{equation}
which is a degenerate scalar-tensor Lagrangian: degenerate in the sense that the scalar field $p$ lacks a dynamical term, the term $V(p)=p\cdot r(p) - f(r)$ being a mere function of $p$. Nevertheless, the scalar field does have a spacetime dynamics, due to the nonminimal coupling with the scalar curvature $R$. The variation of $L_H$ w.r.~to $p$, upon using the identity $f'(r(p))\equiv p$, gives the equation
\begin{equation}
    R=r(p),\label{e1}
\end{equation}
as expected, while the variation w.r.~to $g_{\mu\nu}$ gives
\begin{equation}
    p\left(R_{\mu\nu}-\frac{1}{2}Rg_{\mu\nu}\right)= \nabla_\mu\nabla_\nu p - g_{\mu\nu}\Box p+\frac{1}{2}(f(r)-p\cdot r(p))g_{\mu\nu} .\label{e2}
\end{equation}
\noindent By definition of the function $r(p)$, (\ref{e1}) is equivalent to $p=f'(R)$: plugging this equation into (\ref{e2}), one easily retrieves the fourth order equation (\ref{NLeq}). 

It is worth noting that for $p\ne 0$ a few manipulations (see \cite{MS1}) allow one to recast the full system of second order equations (\ref{e1},\ref{e2}) into the equivalent form
\begin{equation}
\begin{cases}  
R_{\mu\nu}-\frac{1}{2}Rg_{\mu\nu}=p^{-1}\nabla_\mu\nabla_\nu p-\frac{1}{6}\,(p^{-1}f(r)+ r)g_{\mu\nu}\\[10pt]
\Box p = \frac{2}{3}\,f(r)-\frac{1}{3}\,p\cdot r\\
\end{cases}  \label{eeg}
\end{equation}
Therefore, the Legendre transformation indeed allows to replace the fourth order equation (\ref{NLeq}) of a $f(R)$ model with an equivalent second order general-relativistic model (including Einstein equation) with a self-interacting scalar field, where the metric is the \emph{original} metric $g_{\mu\nu}$, without any conformal rescaling. 

The reason to pass to the ``Einstein frame metric'' $\tilde{g}_{\mu\nu}$ (\ref{cr}) is that in (\ref{eeg}) the effective energy momentum tensor for the scalar field $p$, on the r.h.s.~of the Einstein equation, is not the usual one and is seemingly unphysical: it is, in fact, linear in the second derivatives of $p$, so the energy density has indefinite sign. The Einstein frame provides instead a manifestly physical second order picture \cite{MS1}. 

Conformal rescaling is not part of the Legendre transformation: it is an \emph{additional} step needed to obtain an equivalent Einstein-Hilbert Lagrangian instead of the scalar-tensor Lagrangian (\ref{LH2}), analogous to the \emph{Dicke transformation} introduced in 1962 \cite{D62} (the condition $p\ne 0$ has to be imposed to perform the conformal rescaling as well). 

If matter is coupled to the original Lagrangian, however, after the conformal rescaling the matter Lagrangian is multiplied by a power of $p$, as a consequence of the change of the volume density from $\sqrt{|g|}$ to $\sqrt{|\tilde{g}|}$; thus the scalar field $p$ acts as a \emph{variable coupling factor} between ordinary matter and the metric. But even in the equivalent system (\ref{eeg}) for the dynamical  variables $(g_{\mu\nu},p)$, if gravitating matter is added to the $f(R)$ Lagrangian, then the original matter energy-momentum tensor $T_{\mu\nu}$ is multiplied by $p^{-1}$ in the Einstein equation, so the problem persists. This phenomenon, in fact, occurs whenever any Jordan-Brans-Dicke scalar-tensor theory is recast into an Einstein equation (either by simple manipulations or by conformal rescaling). To overcome this problem, one should manage to have matter minimally coupled to the Einstein frame metric: in some cases, this can be achieved by a suitable coupling to the original metric \cite{MS1}.

All the above facts have already been elucidated (in greater detail) in \cite{MS1,M}. The question, therefore, is whether the procedure envisaged in \cite{CMM} provides a \emph{different} way to obtain an \emph{equivalent} formulation of a $f(R)$ model in terms of an Einsten equation with an auxiliary scalar field, not affecting the matter energy-momentum tensor. 

The notation used in  \cite{CMM}, unfortunately, is rather confusing, since the authors use the same symbol $R$ for the scalar curvature of the metric and for the function that we have denoted above by $r(p)$. Instead, they denote the Legendre transform of $L$ in two ways, $W(\psi)$ or $F(\psi,R)$ (they use $\psi$ instead of $p$ to denote the scalar field), which are declared to have the same meaning: this makes it clear that, in accordance with the usual definition of Legendre transform, $R$, when it appears as an argument of $F$, is indeed the function $R(\psi)$ (the inverse of $\psi=f'(R)$), and not the scalar curvature of the metric. 

Then they write: \emph{``the equations obtained by varying the action associated with the Lagrangian $F(\psi, R) = \psi R-f(R)$ with respect to the spacetime metric $g$ are equivalent to the Lagrangian equations obtained by varying the action associated with the Lagrangian $f(R)$''}. Now, this is simply wrong: it is the same as claiming that if one sets the variation of the integral of the Hamiltonian to zero one gets the Hamilton equations, whereas one only gets $\frac{\partial H}{\partial q^\mu}=0$ and $\frac{\partial H}{\partial p_\mu}=0$. The action which is equivalent to the $f(R)$ action would instead be defined by the Helmholtz Lagrangian (\ref{LH2}), i.e.~by the function $\psi R(g)-W(\psi)$. The function $F(\psi, R)\equiv W(\psi)$, the Legendre transform of $L$, does \emph{not} depend on the metric $g_{\mu\nu}$, because $f(R)$ is supposed to depend on $R$ alone; the variation of $W(\psi)$ is nothing but
$$
\delta W(\psi)= \left(R + \psi\frac{dR}{d\psi}-f'(R)\frac{dR}{d\psi}\right)\delta\psi=R(\psi)\delta \psi
$$
(on account of the identity $f'(R(\psi))\equiv \psi$). Imposing $\delta W=0$ would then imply \mbox{$R(\psi)=0$,} an evident nonsense. 

Incidentally, even the metric variation of $[\psi R(g)-f(R(\psi))]\sqrt{|g|}$ would not be the same as the variation of $f(R(g))\sqrt{|g|}$.

Accordingly, in their subsequent example, $f(R)=R+\epsilon R^2$, from the $\psi$-variation of $W(\psi)$ they get $\psi=1$ (i.e.~\mbox{$R=0$}). Actually, this holds provided $\epsilon\ne 0$, otherwise the condition $f''(R)\ne 0$ would be violated, and $R(\psi)$ and $W(\psi)$ would simply not exist; on the contrary, they infer that $\psi=1$ implies $\epsilon= 0$ in the original Lagrangian, and conclude that \emph{``this suggests that the general \emph{metric} variation of the ‘W-action’ \emph{[...]} is somehow related to the Einstein-Hilbert Lagrangian.''} 

As a matter of fact, the density $W(\psi)\sqrt{|g|}$ does not contain derivatives of the dynamical variables -- neither of $\psi$ nor of the metric -- and the variation of its integral cannot generate any differential equations at all. 

The authors of \cite{CMM} then observe that the metric variation of the $W$-action equals the variation of the Brans-Dicke action (without potential) defined by $\psi R(g)$ minus the variation of the original $f(R)$ action: this is true, but only if one assumes $\psi=f'(R)$ to be an identity holding independently. Hence they derive that the metric variation of the $W$-action is proportional to $W$ itself. Actually, the metric variation of $W(\psi)\sqrt{|g|}$ equals $-\frac{1}{2}W(\psi)g_{\mu\nu}\sqrt{|g|}\delta g^{\mu\nu}$ simply because $W(\psi)$ does not depend at all on the metric, so only the volume element is affected. 

Anyway, one should conclude that the extremals of the $W$-action are nothing else than the (constant) solutions $\psi$ of \mbox{$W(\psi)=0$} and \mbox{$R(\psi)=0$}. Instead, in \cite{CMM} the idea of equating the $W$-action to the difference of the BD-action and the $f(R)$ original action leads (without further explanation) to the following ``Einstein–Legendre Lagrangian'': 
\begin{equation}
    L_{EL}=\left[R(g)+F(\psi,R(\psi))+\frac{1}{2}g^{\mu\nu}\nabla_\mu\psi\nabla_\nu\psi\right]\sqrt{|g|}+L_{mat}.
\end{equation}
Here, the authors' ambiguity in the use of $R$ is resolved by their subsequent equation (7), which is only compatible with the above interpretation of the notation (they insert the volume element $\sqrt{|g|}$ in the action integral instead of including it in the definition of $L_{EL}$ and $L_{mat}$, but this is a mere notational choice). 

Now, the variation of $L_{EL}$ w.r.~to the metric gives (in the sequel we revert to denoting $R(\psi)$ by $r$, to avoid misunderstandings)
\begin{equation}
    R_{\mu\nu}-\frac{1}{2}Rg_{\mu\nu}=\frac{1}{2}\big(\psi r -f(r)\big)g_{\mu\nu}-\frac{1}{2}\nabla_\mu\psi\nabla_\nu\psi+\frac{1}{4}\nabla_\lambda\psi\nabla^\lambda\psi g_{\mu\nu}+T^{\mathrm{(mat)}}_{\mu\nu}
\end{equation}
while the variation w.r.~to $\psi$ gives 
\begin{equation}
\Box \psi=r.
\end{equation}

Are these equations equivalent to (\ref{NLeq})? It one takes for granted that \mbox{$p=f'(R(g))$,} then $(g_{\mu\nu},\psi)$ are exactly the same variables $(g_{\mu\nu},p)$ occurring in (\ref{eeg}). Thus, since it has been proven that (\ref{NLeq}) is equivalent to the system (\ref{eeg}), the equations above should be equivalent to (\ref{eeg}) as well.

The trouble is that the variation of the Lagrangian $L_{EL}$ w.r.~to $\psi$ \emph{does not} yield the equation $R(g)=r(\psi)$, so the relation $\psi=f'(R(g))$ cannot be derived from the field equations (in contrast to (\ref{eeg}), where this relation can be recovered by combining the second equation with the trace of the first equation). We only know, by the definition of $r(\psi)$, that $f'(r(\psi))\equiv \psi$. 

If it were true that $R(g)=r(\psi)$ (and equivalently $\psi=f'(R(g))$) on all solutions of the field equations, then one should have $\Box f'(R)=R$, while taking the trace of (\ref{NLeq}) one finds instead $\Box f'(R)=\frac{2}{3}f(R)-\frac{1}{3}f'(R)R+\frac{1}{3}T^{\mathrm{(mat)}}_{\mu\nu}g^{\mu\nu}$. 

For instance, let us check what happens in the case $f(R)=R+\epsilon R^2$ which is considered by the authors in the subsequent section (here we restrict to the vacuum case, for simplicity). In this case one has $r(\psi)=\frac{\psi-1}{2\epsilon}$, and the Einstein-Legendre Lagrangian becomes
\begin{equation}
    L_{EL}=\left[R(g)+\dfrac{(\psi-1)^2}{4\epsilon}+\frac{1}{2}g^{\mu\nu}\nabla_\mu\psi\nabla_\nu\psi\right]\sqrt{|g|};
\end{equation}
hence one finds the equations
\begin{equation}
    R_{\mu\nu}-\frac{1}{2}Rg_{\mu\nu}=\frac{(\psi-1)^2}{8\epsilon}g_{\mu\nu}-\frac{1}{2}\nabla_\mu\psi\nabla_\nu\psi+\frac{1}{4}\nabla_\lambda\psi\nabla^\lambda\psi g_{\mu\nu}\label{uffa1}
\end{equation}
and 
\begin{equation}
\Box \psi=\frac{\psi-1}{2\epsilon}\label{uffa2}.
\end{equation}
Taking the trace of (\ref{uffa1}) one finds
\begin{equation}
R=\frac{1}{4}\nabla_\lambda\psi\nabla^\lambda\psi+\frac{(\psi-1)^2}{2\epsilon}
\end{equation}
which is compatible with $R=\frac{\psi-1}{2\epsilon}$ only if $\psi$ fulfills the first order equation $\nabla_\lambda\psi\nabla^\lambda\psi=\frac{2}{\epsilon}(\psi-1)(\psi-2)$, which does not hold in general for solutions of the field equation (\ref{uffa2}). If, for these particular solutions, one evaluates the trace of (\ref{NLeq}) one gets an algebraic equations for $\psi$: since the only constant solution of (\ref{uffa2}) is $\psi=1$, it turns out that the latter, which is equivalent to $R\equiv 0$, is the \emph{only} solution of (\ref{uffa1}, \ref{uffa2}) for which $\psi=f'(R)$ holds and which is also a solution of (\ref{NLeq}). 

In conclusion, the statement that \emph{``in the `EL representation' the original $f(R)$ theory acquires a particularly simple form given by second order Einstein equations with the scalar field $\psi = f'(R)$ coupled minimally''} appears to be untenable. The ``Einstein-Legendre Lagrangian'' advocated by the authors of \cite{CMM} is not equivalent to the original $f(R)$ Lagrangian and, in general, the equations derived from $L_{EL}$ are incompatible with $\psi = f'(R)$. 

The Lagrangian $L_{EL}$ is related to the original $f(R)$ model only in that the potential of the scalar field $\psi$ is set to coincide with the Legendre transform of the original Lagrangian, but the dynamics produced by the two Lagrangians cannot even be compared with each other, since there is no way to obtain from the variation of $L_{EL}$ a functional relation between $\psi$ and $R$ -- holding for all solutions -- which would be needed to reconstruct a fourth order equation for the metric alone.

\bigskip

Alas, it seems unlikely that anything new can emerge in the realm of metric $f(R)$ theories that has not already been scrutinized in the last decades.

There is, however, one fact to which little attention is paid in the general accounts of the Legendre transformation, and which does not seem to have been exploited in current cosmological applications (with some notable exceptions such as \cite{ext}, which however deal with Palatini $f(R)$ theories). The condition $f''(R)\ne 0$ ensures the local existence of a function $r(p)$ entering the potential of the auxiliary scalar field $p$ in the second order picture of the model: but this function is not unique if $f'(R)$ is not globally invertible (i.e.~monotonic). For instance, if instead of a quadratic Lagrangian one takes $f(R)=\frac{1}{4}R^4-\frac{3}{2}a^2R^2$, then one gets \emph{three} distinct potentials for the field $p$: one for the solutions where $R<-a$, another for the solutions where $|R|<a$ and a third one covering $R>a$. 

In this sense, the second order picture arising form a single $f(R)$ Lagrangian contains (in both the Jordan and the Einstein frame) a gravitating scalar field which may have \emph{different self-interaction potentials for different values of the spacetime curvature} (the \emph{``Legendre sectors''} of the model): this is a situation that does not arise for a scalar field minimally coupled to the Einstein-Hilbert Lagrangian. The transition from one potential to another occurs at a singularity of the Legendre transformation (in the previous example, at $R=\pm a$), where the function $f'(R)$ is nevertheless smooth: therefore a solution of the original fourth order model corresponds to \emph{patching together solutions with different potentials for the scalar field in different regions of spacetime.} It seems that a systematic investigation of the possible applications of this fact in cosmology is still lacking.


\begin{thebibliography}{}
\frenchspacing

\bibitem{CMM}
\textsc{S. Cotsakis, J.P. Mimoso and J. Miritzis,} Eur.Phys.J.C (2023) 83:433.

\bibitem{old} \textsc{P.W. Higgs,} Nuovo Cim. \textbf{11} (1959) 816\\
\textsc{G. Bicknell,} J. Phys. \textbf{A7} (1974) 1061\\
\textsc{P. Teyssandier, P. Tourrenc,} J.Math.Phys. \textbf{24} (1983) 2793\\
\textsc{B. Whitt,} 
Phys.Lett. \textbf{145B} (1984) 176.

\bibitem{Brans} \textsc{C.H. Brans,} Class. Quantum Grav. \textbf{5} (1988), L197.\\
\textsc{M. Ferraris, M. Francaviglia and G. Magnano,} Class. Quantum Grav. \textbf{7} (1990), 261.

\bibitem{Cap} \textsc{S. Capozziello, P. Martin-Moruno and C. Rubano, C}, Physics Letters B, \textbf{689(4-5)} (2010), 117.

\bibitem{K1} \textsc{J. Kijowski,} Gen. Relat. Gravit. \textbf{9}  (1978), 857.

\bibitem{MFF1}
\textsc{G. Magnano, M. Ferraris and M. Francaviglia,} Gen. Rel. Grav. \textbf{19}  (1987), 465.

\bibitem{JK}
\textsc{A. Jakubiec, J. Kijowski,} Gen. Rel. Grav. \textbf{19}  (1987) 719; Phys. Rev. \textbf{D37} (1988) 1406; J. Math. Phys. 30 (1989) 1073; J. Math. Phys. 30 (1989), 2923.

\bibitem{MFF2}
\textsc{G. Magnano, M. Ferraris, M. Francaviglia,}
Class. Quant. Grav. \textbf{7} (1990), 557.

\bibitem{MFF3}
\textsc{G. Magnano, M. Ferraris, M. Francaviglia,}
J. Math. Phys. 31(1990), 378.

\bibitem{M} \textsc{G. Magnano,} Intern. J. Geom. Meth. in Modern Phys. \textbf{13(08)} (2016), 1640006.

\bibitem{K2}
\textsc{J. Kijowski,} Intern. J. Geom. Meth. Phys. \textbf{13} (2016), 1640008.

\bibitem{MS2}
\textsc{G. Magnano, L.M. Soko\l{}owski,}
Ann. Phys. (N.Y.) \textbf{306} (2003) 1.

\bibitem{MS1}
\textsc{G. Magnano, L.M. Soko\l{}owski,}
Phys. Rev. \textbf{D50} (1994) 5039.

\bibitem{D62}
\textsc{R.H. Dicke,} Phys. Rev. \textbf{125} (1962) 2163

\bibitem{He}
\textsc{H. von Helmholtz,}
\textit{Wissenschaftliche Abhandlungen, Dritter Band}, A. Barth (Leipzig 1895), 203.

\bibitem{ext}
\textsc{P. Pinto, L. Del Vecchio, L. Fatibene and M. Ferraris,}
J. Cosmology and Astroparticle Physics, \textbf{11} (2018), 044.\\
\textsc{L. Del Vecchio, L. Fatibene, S. Capozziello, M. Ferraris, P. Pinto and S. Camera,}
Eur. Phys. J. Plus \textbf{134(5)} (2019)

\end{thebibliography}
\end{document}